# PRESSURE EFFECT AND SPECIFIC HEAT OF RBa$_2$Cu$_3$O$_x$ AT DISTINCT CHARGE CARRIER CONCENTRATIONS: POSSIBLE INFLUENCE OF STRIPES

S. I. SCHLACHTER[1], U. TUTSCH[1], W. H. FIETZ[1], K.-P. WEISS[1], H. LEIBROCK[1], K. GRUBE[1], Th. WOLF[1], B. OBST[1], P. SCHWEISS[2], AND H. WÜHL[1,3].

*Forschungszentrum Karlsruhe, [1]ITP and [2]IFP, 76021 Karlsruhe, Germany.*
*[3]Universität Karlsruhe, IEKP, 76128 Karlsruhe, Germany.*

In YBa$_2$Cu$_3$O$_x$, distinct features are found in the pressure dependence of the transition temperature, d$T_c$/d$p$, and in $\Delta C_p \cdot T_c$, where $\Delta C_p$ is the jump in the specific heat at $T_c$: d$T_c$/d$p$ becomes zero when $\Delta C_p \cdot T_c$ is maximal, whereas d$T_c$/d$p$ has a peak at lower oxygen contents where $\Delta C_p \cdot T_c$ vanishes. Substituting Nd for Y and doping with Ca leads to a shift of these specific oxygen contents, since oxygen order and hole doping by Ca influences the hole content $n_h$ in the CuO$_2$ planes. Calculating $n_h$ from the parabolic $T_c(n_h)$ behavior, the features coalesce for all samples at $n_h \approx 0.11$ and $n_h \approx 0.175$, irrespective of substitution and doping. Hence, this behavior seems to reflect an intrinsic property of the CuO$_2$ planes. Analyzing our results we obtain different mechanisms in three doping regions: $T_c$ changes in the optimally doped and overdoped region are mainly caused by charge transfer. In the slightly underdoped region an increasing contribution to d$T_c$/d$p$ is obtained when well ordered CuO chain fragments serve as pinning centers for stripes. This behavior is supported by our results on Zn doped NdBa$_2$Cu$_3$O$_x$ and is responsible for the well known d$T_c$/d$p$ peak observed in YBa$_2$Cu$_3$O$_x$ at $x \approx 6.7$. Going to a hole content below $n_h \approx 0.11$ our results point to a crossover from an underdoped superconductor to a doped antiferromagnet, changing completely the physics of these materials.

## 1 Introduction

In the past years a lot of work has been done to investigate the $T(n_h)$ phase diagram of cuprate superconductors, where $n_h$ denotes the hole concentration in the CuO$_2$ planes. For $n_h = 0$ the cuprates are Mott insulators with long-range antiferromagnetic order. With increasing $n_h$ this antiferromagnetic order is destroyed rapidly and superconductivity sets in. The superconducting transition temperature $T_c$ grows with $n_h$ in the underdoped region up to a maximum transition temperature $T_{c,max}$ at optimum doping $n_{h,opt}$ and then decreases again in the overdoped region.

There is growing evidence that some physical peculiarities in the underdoped region and possibly the superconductivity are a result of a spin-charge separation in the CuO$_2$ planes into line-shaped spin- and hole-rich regions. This so-called stripe phase seems to vanish beyond $n_{h,opt}$ where magnetic correlations become negligible. The observed stripes at 1/8 doping ($n_h = 0.125$) in the La214 compound show that stripe mobility has crucial influence on superconductivity [1]. The introduction of Zn in the CuO$_2$ planes, for example, suppresses superconductivity most likely due to the pinning of fluctuating stripes [2].

In the RBa$_2$Cu$_3$O$_x$ system (R123, R= rare earth element) the whole underdoped, the optimally doped and the slightly overdoped region can be investigated by varying the oxygen content $x$. With partial substitution of Ca$^{2+}$ for Y$^{3+}$ experiments can even be extended to the heavily overdoped region. In contrast to the La214 compounds the R123 system contains also CuO chains serving as a charge reservoir. This allows changes of the hole concentration of a sample via pressure application without changing the chemistry of the sample, because application of pressure leads to charge transfer from the CuO chains to the CuO$_2$ planes due to different length changes of hard and soft bonds [3]. In this work we present measurements of $T_c$, the hydrostatic pressure effect $dT_c/dp$ and the specific heat of Ca doped Y123 and Zn doped Nd123 single crystals.

## 2 Experimental

In order to determine the pressure dependence of $T_c(p)$ we performed ac-susceptibility measurements under absolutely hydrostatic pressure conditions ($p \leq 0.6$ GPa) with He gas as pressure-transmitting medium Pressure-induced oxygen-ordering effects [4], leading to the creation of additional holes in the CuO chains and therefore as well to an increase of the hole concentration in the $CuO_2$ planes, have been avoided by exposing the samples to high pressures only at temperatures below 110 K.

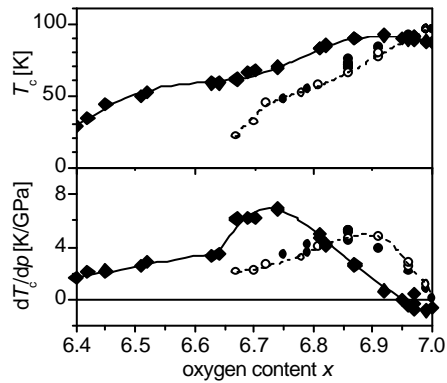

Fig. 1: $T_c$ and pressure effect of Nd123 (?) and $(Nd_{0.93}Y_{0.07})123$ (?) in comparison with $T_c$ and pressure effect of Y123 (?).

The specific-heat measurements were performed by a continuous heating technique with samples of about 30 mg. To substract the normal state background from the specific heat we used data from a Nd/Ba substituted sample with a strong depressed superconducting transition. A mean-field type specific-heat jump $\Delta C_p/T_c$ at $T_c$ has been obtained, analyzing the superconducting transition with an entropy conserving construction.

The $Y_{1-y}Ca_yBa_2Cu_3O_x$, as well as the $NdBa_2(Cu_{1-z}Zn_z)_3O_x$ samples were grown in Y stabilized $ZrO_2$ crucibles. EDX analysis showed that in the Nd123 samples, corrosion of the crucibles during the growth process lead to a small occupation of about 7 at% Y on Nd sites. Since, however, $T_c$ and pressure effect of these samples and of Y free Nd123 samples grown in $BaZrO_3$ crucibles show only very small differences compared to $T_c$ and pressure effect of pure Y123 (Fig. 1), in the following we will not distinguish between Nd123 with and without Y impurities. EDX analysis and neutron diffraction studies gave no hints for other impurities or Nd/Ba misoccupations.

The oxygen contents $x(T,p)$ of our samples were adjusted by annealing the samples under flowing oxygen, oxygen/argon or oxygen/nitrogen mixtures. For the Y123 and Nd123 samples the appropriate temperature and oxygen partial pressure for a distinct oxygen content were deduced from Refs. [5] and [6]. In $Y_{1-y}Ca_yBa_2Cu_3O_x$ (YCa123) the oxygen content is reduced in comparison to the expected values from Refs. [5] and [6] by approximately $y/2$ due to the different valence of $Y^{3+}$ and $Ca^{2+}$ [7].

## 3 Results and Discussions

For many cuprate superconductors, including $La_{2-x}Sr_xCuO_4$, $La_{2-x}Sr_xCaCu_2O_6$ and $Y_{1-y}Ca_yBa_2Cu_3O_x$ with various Ca and oxygen contents, $T_c(n_h)$ follows a universal parabolic behavior [8]. The optimum hole concentration $n_{h,opt} \approx 0.16$ is common for all these superconductors, whereas the maximal achievable $T_c$ depends on the particular system.

Following this dependence we measured for each sample the $T_c$ values at various oxygen contents to obtain $T_{c,max}$. With these values we determined $n_h$ for each single crystal at the particular oxygen content. In Fig. 2, these $n_h$ values are plotted versus oxygen content. For the pure Y123 we find at an oxygen content of $6.5 < x < 6.65$ the well known 60 K plateau which is caused by oxygen ordering. Due to the larger lattice parameters of Nd123, oxygen ordering is diminished. Therefore, the doping efficiency of oxygen is also diminished and the same $n_h$ values require higher oxygen contents than in Y123 [9]. Due to the doping effect of Ca, the Ca doped Y123 samples show much higher hole concentrations than Y123 at the same oxygen content. The missing signature of the 60 K plateau is caused by the decreasing tendency to a well-ordered oxygen sublattice with increasing Ca content.

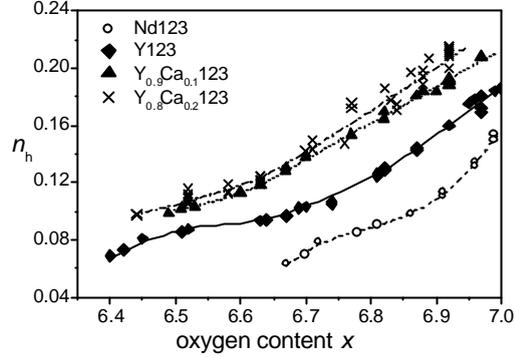

Fig. 2: Hole concentration determined from $T_c$, $T_{c,max}$ and the parabolic $T_c(n_h)$ behavior [8] versus oxygen content.

When our results are plotted as a function of the hole content, however, despite the different doping mechanisms, the $dT_c/dp(n_h)$ dependence of the different systems look quite similar, as well as the $\Delta C_p \cdot T_c(n_h)$ dependence (Fig. 3). In the underdoped region the pressure effects of the different systems peak at $n_h \approx 0.11$, where $\Delta C_p \cdot T_c$ vanishes. On the other hand, in the overdoped region the pressure effects of all samples show an almost linear behavior crossing zero at $n_h \approx 0.175$. Exactly at this doping level, $\Delta C_p \cdot T_c$, which is a measure of the condensation energy, shows a maximum for all samples (except for Y123, where a further increase is visible because the well ordered CuO chains become superconducting). The maxima in the condensation energies and the zero pressure effects together with other experimental results in the literature [10] suggest that superconductivity in the slightly overdoped region is extremly stable.

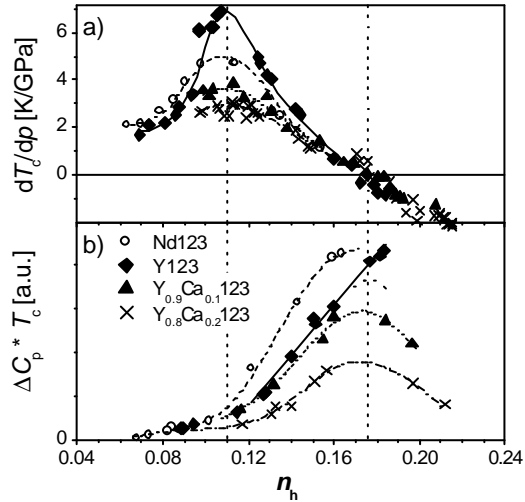

Fig. 3: a) Pressure effect $dT_c/dp$ and b) $\Delta C_p \cdot T_c$ versus hole concentration in the CuO$_2$ planes.

The almost linear decrease of the hydrostatic pressure effect in the overdoped region can be understood in terms of pressure-induced charge-transfer [11] from the CuO chains to the CuO$_2$ planes, which is mainly caused by pressure along the $c$-axis direction. For pressure along the $a$- and $b$-axis direction pressure-induced charge-transfer can be neglected [12]. According to the parabolic $T_c(n_h)$ behavior, a constant pressure-induced charge-transfer rate $dn_h/dp$ leads to a linear $dT_c/dp(n_h)$ behavior with positive pressure effects in the underdoped, a zero pressure effect in the optimally doped and negative pressure effects in the overdoped region. Exactly this behavior was found for $n_h > 0.11$ for uniaxial $c$-axis pressure, determined either by direct measurements [13] or by thermal expansion measurements via Ehrenfest's theorem [14-16]. Data from these investigations are shown in the inset of Fig. 4. The solid line for $n_h > 0.11$ in the inset of Fig. 4 is the derivative of the $T_c(n_h)$ parabola with $dn_h/dp \approx 3.7 \cdot 10^{-3}$ GPa$^{-1}$ [17]. In this doping regime, $T_c$ changes by uniaxial $c$-axis pressure are therefore mainly caused by charge transfer. Due to this simple behavior for $n_h > 0.11$ we can divide off the effect of charge-transfer by subtracting the solid line in the inset of Fig. 4 from our hydrostatic $T_c(p)$ data. As a result we obtain a measure for the in-plane compression, namely the sum of $a$- and $b$-axis pressure effect [18].

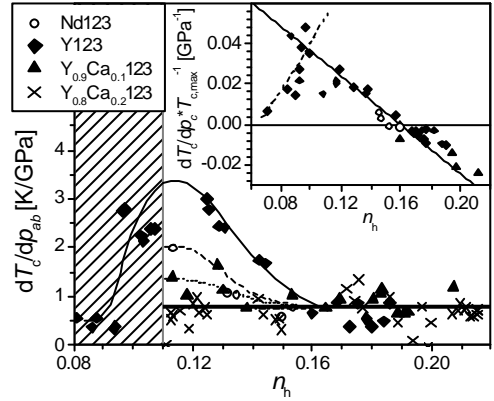

Fig. 4: Non-charge-transfer pressure effect $dT_c/dp_{ab}$. Inset: $dT_c/dp_c \cdot T_{cmax}^{-1}$ versus $n_h$ for Y123 (?, Refs. [13, 14]), YCa123 (?, Refs. [15, 16]) and Nd123 (?, Ref. [16]). The solid line sketches the pure charge-transfer pressure effect calculated for $dn_h/dp \sim 3.7 \cdot 10^{-3}$ GPa$^{-1}$ [17].

In Fig. 4 the effect of in-plane compression on $T_c$, $dT_c/dp_{ab}$, shows nearly constant values in the overdoped and optimally doped regime. For pure Y123 $dT_c/dp_{ab}$ rises in the underdoped region with decreasing doping down to $n_h > 0.11$. For Nd123 this effect is smaller and almost absent for the Ca doped samples. The fact that this behavior is not correlated to charge transfer is confirmed by the quite similar looking pressure effect of $La_{2-x}Ba_xCuO_4$ where charge-transfer effects are known to be absent [19].

This coincidence points to a possible explanation for the $T_c(p)$ peak at $n_h \approx 0.11$. For $La_{2-x}Ba_xCuO_4$ the drastic collapse of $T_c$ at $n_h = 1/8$ was interpreted to be caused by stripe pinning. Under pressure these stripes are depinned and $T_c$ recovers [20].

For fully oxygenated R123 we find no argument for the existence of stripe pinning. But with a reduction of the oxygen content we have in Y123 well ordered CuO chains in an oxygen depleted neighborhood and such a configuration may pin stripes. This idea would also explain the different $T_c(p)$ changes under uniaxial $a$- or $b$-axis pressure [14, 15] because a compression perpendicular or parallel to the stripe direction would naturally cause different effects on stripe pinning. An analysis of the experimental data shown in Fig. 4 supports this idea. With decreasing hole content we find a drastic $dT_c/dp$ increase for Y123. For Nd123 with a much lower tendency to oxygen ordering this increase is dramatically diminished and for Ca doped samples with a large disorder in the oxygen sublattice we have almost no peak effect in $dT_c/dp$.

Another striking feature of an in-plane compression on $T_c$ of the different R123 systems with quite different structural parameters is that they show a sharp decrease of $dT_c/dp$ below $n_h \approx 0.11$. At the same hole concentration also the $c$-axis pressure effect (that is attributed to pressure-induced charge-transfer) rapidly decreases. In addition we find other physical properties of different cuprate superconductors showing peculiarities at this hole content. The copper isotope effect $d\ln(T_c)/d\ln(m^{Cu})$ of $YBa_2Cu_3O_x$ [21] and the oxygen isotope effects $d\ln(T_c)/d\ln(m^O)$ of $La_{2-x}Ba_xCuO_4$ [22] and $La_{2-x}Sr_xCuO_4$ [23] show drastic changes at $n_h \approx 0.11$, quite similar to the doping dependence of the thermal resistivity of $YBa_2Cu_3O_x$ [24]. At $n_h \approx 0.11$ also the doping dependence of the room-temperature thermopower of many cuprates changes from an exponential to a linear behavior [25]. Additionally, in $La_{1.6-x}Nd_{0.4}Sr_xCuO_4$ at $x = n_h \approx 0.11$ the doping dependence of the magnetic incommensurability $e$, which is a measure for the distance of charge stripes, changes from $e = x$ to $e \approx$ constant [26]. Hunt et al. [27] showed that below a hole content $n_h \approx 0.11$ the amount of static stripes is increased drastically. From these arguments we conclude, that below $n_h \approx 0.11$ the physics in these materials is changing – above $n_h \approx 0.11$ we deal with a doped superconductor but below $n_h \approx 0.11$ we have to look at a doped antiferromagnet.

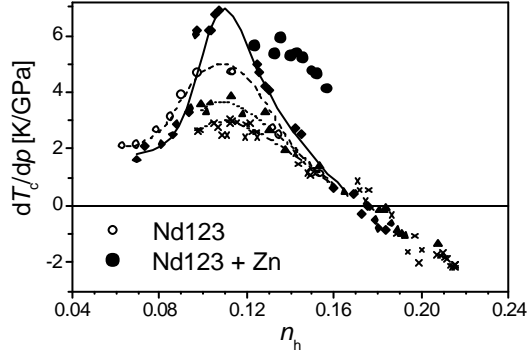

Fig. 5: Pressure effect of Zn doped Nd123 (?) in comparison to the pressure effects of Zn free R123 samples.

In addition to the Ca doped R123 samples, we also investigated $NdBa_2(Cu_{0.98}Zn_{0.02})_3O_x$ single crystals. Zn is known to substitute for Cu in the $CuO_2$ planes and to depress $T_c$ by pair-breaking effects. Therefore, $n_h$ could not be calculated from the parabolic $T_c(n_h)$ behavior. Assuming that Zn doping does influence neither the oxygen content achieved under certain annealing conditions nor the hole concentration in the $CuO_2$ planes, we estimated $n_h$ from the tempering conditions and the $n_h(x)$ dependence of Zn free samples shown in Fig. 2. This assumption was confirmed for Zn-doped YCa123 [28].

In Refs. [29] and [30] it was shown that around the Zn impurities small non-superconducting domains exist. Such non-superconducting domains may serve as pinning centers for stripes and would be independent of the oxygen content. In the pressure induced depinning picture one would then expect large pressure effects not only for underdoped samples but for fully oxygenated material, too. Fig. 5 shows the doping dependence of the pressure effects of our Zn doped samples in comparison to the pressure effects of the other R123 samples. At approximately optimum doping, where the pressure effects of the other samples are about 0.8 K/GPa the pressure effect of the Zn doped Nd123 samples is approximately 5 times larger. With decreasing hole concentration in the $CuO_2$ planes the pressure effect even increases to higher values than the maximum pressure effect of the Zn free Nd123 samples beeing consistent with the idea of pressure-induced depinning of stripes.

## 4   Conclusions

We measured $T_c$, pressure effect $dT_c/dp$ and the specific heat of various Zn and Ca doped R123 single crystals and found two distinct charge carrier concentrations in the underdoped and overdoped region, where $dT_c/dp$ and $\Delta C_p \cdot T_c$ show distinct features. In the overdoped region around $n_h \approx 0.175$ superconductivity is very stable and $T_c$ changes under pressure are mainly caused by charge transfer. In the underdoped region the large pressure effects, which are not related to charge transfer, can be attributed to depinning of charged stripes. This idea is confirmed by the large pressure effect of Zn doped Nd123 samples even in the nearly optimally doped region. The breakdown of the pressure effects at $n_h < 0.11$ with decreasing hole concentration is assigned to the crossover from a doped superconductor to a doped antiferromagnet.

**References**

1. S.A. Kivelson *et al.*; *Nature* **393**, 550 (1998).
2. Y. Koike *et al.*; Proceedings of the 2000 international workshop on superconductivity, June 19-22, 2000; Shimane, Japan.
3. J.D. Jorgensen *et al.*; *Physica* **C 171**, 93 (1991).
4. R. Sieburger and J.S. Schilling; *Physica* **C 173**, 403 (1991). R. Benischke *et al.*; *Physica* **C 203**, 293 (1992); W.H. Fietz *et al.*; *Physica* **C 270**, 258 (1996). V.G. Tissen *et al.*; *Physica* **C 316**, 21 (1999).
5. T.B. Lindemer *et al.*; *J. Am. Ceram. Soc*. **72**, 1775 (1989).
6. T.B. Lindemer *et al.*; *Physica* **C 255**, 65 (1995).
7. B. Fisher *et al.*; *Phys. Rev*. **B 47**, 6054 (1993); C. Glédel *et al.*; *Physica* **C 165**, 437 (1990).
8. M.R. Presland *et al.*; *Physica* **C 176**, 95 (1991); J.L. Tallon *et al.*; *Phys. Rev*. **B 51**, 12911 (1995).
9. U. Tutsch *et al.*; *J. of Low Temp. Physics* **117**, 951 (1999).
10. J.L. Tallon and J.W. Loram; cond-mat/0005063.
11. J.D. Jorgensen *et al.*; *Physica* **C 171**, 93 (1990).
12. H.A. Ludwig *et al.*; *Physica* **C 197**, 113 (1992).
13. H.A. Ludwig, PhD Thesis, University of Karlsruhe (1998), *FZKA* **6117;** U. Welp *et al.*; *J. of Supercond*. **7**, 159 (1994); U. Welp *et al.*; *Phys. Rev. Lett*. **69**, 2130 (1992).
14. O. Kraut *et al.*; *Physica* **C 205**, 139 (1993).
15. C. Meingast *et al.*; *J. of Low Temp. Phys*. **105**, 1391 (1996).
16. V. Pasler, PhD Thesis, University of Karlsruhe (1999), *FZKA* **6415**.
17. S.I. Schlachter *et al., Physica* **C 328**, 1 (1999).
18. W.H. Fietz *et al*.; to be published in *Physica* **C**.
19. W.J. Liverman *et al*.; *Phys. Rev*. **B 45**, 4897 (1992).
20. J.S. Zhou and J.B. Goodenough; *Phys. Rev*. **B 56**, 6288 (1997).
21. J.P. Franck and D.D. Lawrie; *J. of Low Temp. Phys*. **105**, 801 (1996).
22. M.K. Crawford *et al*.; *Phys. Rev*. **B 41**, 282 (1990).
23. G. Zhao *et al*.; *J. Phys.: Cond. Mat*. **10**, 9055 (1998).
24. J.L. Cohn *et al*.; *Phys. Rev*. **B 59**, 3823 (1999).
25. S.D. Obertelli *et al*.; *Phys. Rev*. **B 46**, 14928 (1992).
26. J.M. Tranquada *et al*.; *Phys. Rev. Lett*. **78**, 338 (1997).
27. A.W. Hunt *et al*.; *Phys. Rev. Lett*. **82**, 4300 (1999).
28. J.L. Tallon *et al*.; *Phys. Rev. Lett*. **75**, 4114 (1995).
29. B. Nachumi *et al*.; *Phys. Rev. Lett*. **77**, 5421 (1996).
30. S.H. Pan *et al*.; *Nature* **403**, 746 (2000).